\documentclass[universe,article,accept,pdftex,moreauthors]{Definitions/mdpi} 
\makeatletter      
                    \renewcommand{\maketag@@@}[1]{\hbox{\m@th\normalsize\normalfont#1}}%
                    \makeatother
\firstpage{1} 
\pubvolume{1}
\issuenum{1}
\articlenumber{0}
\pubyear{2024}
\copyrightyear{2024}
\externaleditor{Academic Editor: Firstname \\Lastname}
\datereceived{29 December 2023} 
\daterevised{30 April 2024} 
\dateaccepted{2 May 2024} 
\datepublished{ } 
\hreflink{https://doi.org/} 

\usepackage{amssymb, graphics}
\usepackage{pgfplots}
\usepackage{textcomp}
\usepackage{subcaption}
\usepackage{indentfirst}
\usepackage{amsmath}
\usepackage{commath}
\usepackage{etoolbox}
\makeatletter
\usepackage{graphicx,bm}
\usepackage{float}
\usepackage{bbold}
\usepackage{verbatim}
\usepackage[utf8]{inputenc}

\Title{Dip-Bump Structure in Proton's Single Diffractive Dissociation at the Large Hadron Collider} 

\TitleCitation{Dip-Bump Structure in Proton's Single Diffractive Dissociation at the Large Hadron Collider}

\Author{L\'aszl\'o Jenkovszky 
 $^{1,}$*, Rainer Schicker $^{2}$ and Istv\'an Szanyi $^{3,4,5}$} 
\AuthorNames{L\'aszl\'o Jenkovszky, Rainer Schicker and Istv\'an Szanyi}

\AuthorCitation{Jenkovszky, L.; Schicker, R.; Szanyi, I.}

\address{%
$^{1}$ \quad Bogolyubov Institute for Theoretical Physics (BITP), 
Ukrainian National Academy of Sciences,  14-b, Metrologicheskaya str.,
03680  Kiev,  Ukraine\\
$^{2}$ \quad Physikalisches Institut, University Heidelberg, Im Neuenheimer Feld 226,
69120 Heidelberg, Germany; schicker@physi.uni-heidelberg.de\\
$^{3}$ \quad Department of Atomic Physics, E\"otv\"os University,
P\'azm\'any P. s. 1/A,  H-1117 Budapest, Hungary; szanyi.istvan@wigner.hu\\
$^{4}$ \quad Wigner Research Centre for Physics, 114, P.O. Box 49,  H-1525 Budapest, Hungary\\
$^{5}$ \quad MATE Institute of Technology,  K\'aroly R\'obert Campus, M\'atrai \'ut 36, H-3200 Gy\"ongy\"os, Hungary}

\corres{Correspondence:  jenk@bitp.kiev.ua}

\abstract{By extending the dipole Pomeron (DP) model, successful in describing elastic nucleon--nucleon scattering, to proton single diffractive dissociation (SD), we predict a dip-bump structure in the squared four-momentum transfer ($t$) distribution of proton's SD. Structures in the $t$ distribution of single diffractive dissociation are predicted around $t=-4$ GeV$^2$ at LHC energies in the range of 3 GeV$^2$ $\lesssim|t|\lesssim$ 7 GeV$^2$. 
Apart from the dependence on $s$ (total energy squared) and $t$ (squared momentum transfer), we predict also a dependence on missing masses.  
We include the minimum set of Regge trajectories, namely the Pomeron and the Odderon, indispensable at the LHC. Further generalization, e.g., by the inclusion of non-leading Regge trajectories, is straightforward.  
The present model contains two types of Regge trajectories: those connected with $t-$channel exchanges (the Pomeron, the Odderon, and non-leading (secondary) reggeons) appearing at small and moderate $-t$, where they are real and nearly linear, as well as direct-channel trajectories $\alpha(M^2)$ related to missing masses. In this paper, we concentrate on structures in $t$ neglecting (for the time being) resonances in~$M^2$.}


\begin{document}

\section{Introduction}
High-energy hadron diffraction has been studied experimentally and theoretically by many groups of authors. Apart from details, there is consensus concerning global features such as the rise with energy or the shrinkage of the diffraction cone, etc. At the same time, many important details are still either unknown or disputable, such as the production of resonances in missing mass and possible structure in the differential cross section of diffractive dissociation, similar to the well-known dip-bump in high-energy proton--proton scattering. In the present paper we suggest a model predicting such structures in single dissociation. We give numerical predictions to be tested in future experiments. As far as we know, this is the first prediction of this kind.

The majority of papers on high-energy particle physics start and end with a phrase stating that the right theory is the standard model, with quantum  chromodynamics (QCD) being its foundation. However attractive, this is not true until the problem of confinement is solved. Examples are hadron scattering in the ``soft'' (low and moderate $t$) region, considered in this paper, or hadron spectroscopy or the nature of  confining forces producing  heavy  glueballs lying on the Pomeron or Odderon trajectory from massless gluons.  
   
Our goal is to study the elastic and inelastic diffractive scattering of hadrons (protons and antiprotons) including resonance production in these reactions. We rely on methods of the analytic $S$ matrix theory, including the Regge-pole model appended and extended by analyticity, duality and unitarity, fixing the available freedom and flexibility of the Regge-pole model.

Most of the existing criticism of the Regge-pole approach uses the epithet ``old fashioned'' in contrast with the ``advanced'' standard model (``antique'' vs. ``modern''), implying the uniqueness of the QCD Lagrangian compared to the flexibility of the Regge approach. Let us remind readers that 
the available freedom and flexibility of the Regge approach can be used for its perfection. We prefer to refer to local field theory and analytic $S$ matrix theory in comparing these two possible approaches. We use maximal analyticity, crossing, duality and unitarity  in constructing the $S$ matrix (scattering amplitude) to restrict the available freedom, e.g., in the Regge-pole approach. 

Specifically, we use double poles \cite{Bialkowski, Phillips} instead of simple poles with linear Regge trajectories in the $t-$channel at small $|t|$ as input \cite{Barone, DL}. Otherwise, the trajectories are non-linear complex functions, their form being strongly constrained, as shown in \cite{Trush} and papers cited therein. Without going into details here, we only mention that $t-$channel unitarity requires the presence of trajectory threshold singularities, with the lowest (in the Pomeron trajectory) at $4m_\pi^2$, while at large
$|t|$, the trajectories flatten, resulting in the ``hardening'' of dynamics and a transition from an exponential to a power decrease in the amplitude (or cross section). This is important for the direct channel trajectory $\alpha(M^2).$ 

A typical trajectory featuring the transition from nearly forward to large $|t|$ behaviour~is
\begin{eqnarray}
\alpha(t)=\alpha_0-\sum_i\ln\Biggl(1+\beta_i\sqrt{t_i-t}\Biggr).
\end{eqnarray}

The trajectories in this  duality approach play the role of somewhat dynamical variables. 

In the present paper, we use a dipole Pomeron as the input, extended by a dip-bump mechanism and non-linear trajectories whose form is constrained \cite{Trush} by unitarity and analyticity. Apart from the Pomeron, the Odderon trajectory, required by the data, is included. Lower-lying, non-leading trajectories, importantly below the LHC energy range, are ignored here for simplicity when analyzing elastic scattering. 
 
The non-linearity of Regge trajectories is an essential premise of our approach. The shape of the complex, non-linear trajectories and their observable consequences will be discussed.          

High-energy hadron scattering is characterized by a forward peak followed by possible dips and bumps. The forward peak shrinks with energy (its slope increases), and this is related to the radii of the scattering particles. Multiple dips and bumps may appear in scattering of nuclei, but only a single structure was seen in  $pp$ scattering \cite{Nagy}.  

Proton elastic scattering and diffractive dissociation, single (SD) and double (DD), are closely related reactions. They were studied at the ISR, SPS and FNAL and are being studied at the LHC. Until now, no structures were seen in the differential cross sections of SD or DD.

For SD and DD, we use the model developed in a number of papers \cite{Kuprash, Entropy}. The scattering amplitude in those papers is similar to that of elastic scattering , just as the elastic vertices are replaced by DIS structure functions. 

The position of the structures depends on missing masses as well as on the slopes of the SD and DD cones. The general trend is that the decreasing slope pushes the structures towards larger $-t$. Most of the present measurements at the LHC are in the region of large missing masses in a wide range of $t$. With the present paper, we encourage experimentalists to measure the $-t$ dependence  of SD and DD beyond several GeV$^2$ for different diffrative excitation values. 

\section{Two Components}\label{subsec:Duality}
Any hadron--hadron total cross section (or scattering amplitude) is a sum of two contributions: diffractive and non-diffractive (see Table~\ref{tab:t1}). According to the concept of two-component duality \cite{Barone},
the diffractive component, the smooth background at low energies (here: missing masses) is dual to a Pomeron $t-$channel exchange at high energies, while the non-diffractive
component contains direct channel resonances, dual to high-energy $t-$channel (sub-leading) Reggeon exchanges.

According to our present knowledge about two-body hadron reactions, two distinct classes of reaction mechanisms exist.


The first one includes the formation of resonances in the $s-$channel
and the exchange of particles, resonances or Regge
trajectories in the $t-$channel. The low-energy resonance
behaviour and the high-energy Regge asymptotics are related by
duality, which, at the Born level, or,
alternatively, for tree diagrams can be mathematically formalized
in the Veneziano model, which is a combination of Euler
Beta-functions \cite{Veneziano}.

The second class  does not exhibit resonances at low
energies and its high-energy behaviour is governed by the exchange
of a vacuum Regge trajectory, the Pomeron, with an intercept equal
to or slightly greater than one. Harari and Rosner 
hypothesized that the low-energy non-resonating background is dual
to the high-energy Pomeron exchange, or diffraction. In other
words, the low energy background should extrapolate to high-energy
diffraction in the same way that the sum of narrow resonances sum up
to produce Regge behaviour. However, contrary to the case of
narrow resonances, the Veneziano amplitude, by construction,
cannot be applied to (infinitely) broad resonances. This becomes
possible in a generalization of narrow resonance dual models
called Dual Amplitudes with Mandelstam Analyticity (DAMA),
allowing for (infinitely) broad resonances; see \cite{DAMA} and the references~therein.

In the resonance region, roughly $1\lesssim M\lesssim 4$~GeV, the non-diffractive component of the amplitude is adequately described by a ``reggeized Breit-Wigner'' term , following from the low-energy decomposition of a dual amplitude, with a direct-channel baryon (nucleon, in our case) trajectory with relevant nucleon resonances lying on it.


 \begin{table}[H]
 \caption{Two-component
 duality.\label{tab:t1}}
 \setlength{\tabcolsep}{3.6mm}
 \resizebox{\linewidth}{!}{\begin{tabular}{ccc}
   \toprule
   \boldmath{${\cal I}m A(a+b\rightarrow c+d)$}& \textbf{secondary reggeons}& \textbf{Pomeron} \\
   \midrule
   $s-$channel & $\sum A_{Res}$  & Non-resonant background \\
   \midrule
   $t-$channel & $\sum A_{Regge}$ & Pomeron $(I=S=B=0;\ C=+1)$ \\
   \midrule
   High energy dependence & $s^{\alpha-1},\ \alpha<1$ & $s^{\alpha-1},\ \alpha\geq 1$ \\
   \bottomrule
 \end{tabular}}

 \end{table}

By duality, a proper sum of direct channel resonances produces smooth Regge behaviour, and, to avoid ``double counting'', one should not add the two. Actually, this is true only for an infinite number of resonance poles. For technical reasons, we include only a finite number of resonance poles; moreover, apart from the ``regular'' contribution of the nucleon resonances, we only include poles lying on the $N^*$ trajectory (see \cite{Entropy}). 

The second, diffractive component is essentially the contribution from a Pomeron pole exchange
$\sigma^{Pp}_T\sim (M^2)^{\alpha(0)-1}$, with $\alpha(0)\approx 1.08$ \cite{DL},
and as shown by Donnachie and Landshoff \cite{DL}, this term also can give some contribution to the low-energy (here, missing mass) flat background.

\section{Analyticity, Unitarity, Crossing and Duality}
The present model is inspired by the main ingredients of the analytic $S$ matrix theory, accumulated in dual amplitudes with Mandelstam analyticity (DAMA). DAMA is a flexible representation valid in a wide kinematic range, with ``soft'' and ``hard'' modes of elastic and inelastic hadron scattering. 
It is flexible enough to enable various applications, in particular elastic and inelastic diffraction, the subject of the present paper. To make this point clear, we start by introducing the main ingredients of the approach, namely Regge-poles and non-linear complex Regge trajectories, playing in dual models the role of somewhat dynamical variables. Note that a single-variable Regge trajectory, $\alpha(t)$ connects the scattering region, $t<0$ with particle spectroscopy, $t>0$, thus anticipating duality. 

In this paper, we use the standard Mandelstam relativistic invariant kinematical variables: $s=(p_1+p_2)^2$ is the squared incoming energy, and $t=(p_1-p_3)^2$ is the squared momentum transfer, where $p_1$ and $p_1$ are the four-momenta of the incoming hadrons and $p_3$ is the four-momentum of one of the outgoing hadrons. 

\subsection{Regge Poles and Trajectories}
The formal quantum-mechanical derivation of Regge-poles (Tullio Regge) was preceded by the important empirical observation by G. Chew and S. Frauchi that the hadron scattering amplitude does not factorize in $s$ and $t$. This is supported by the behaviour of the slope of the forward peak, defined as 
\begin{eqnarray}
B(s)=\frac{d}{dt}\log\Bigl(\frac{d\sigma}{dt}(s,t)\Bigr)\bigg|_{t=0},
\end{eqnarray}
where 
\begin{eqnarray}
\frac{d\sigma}{dt}(s,t)\sim |A(s,t)|^2 
\end{eqnarray}
and $A(s,t)$ is the scattering amplitude. 

In most of the papers on the subject, the trajectories are assumed to be real, linear functions. The origins of this prejudice are as follows: (1) the spectrum of most of the observed resonances is nearly compatible with linear real parts of the trajectories; (2) the Veneziano model  works only for linear trajectories; (3) simplicity.  

To complete this brief introduction to Regge trajectories, we remind that they differ by quantum numbers and by their parameters. 

There are boson and fermion trajectories associated with known resonances.  Typical (simplified, approximate) examples are meson
$$
\alpha_{\rho}\approx0.5+0.9t\ (\rho,\ \omega,\ A_2,\ f,...),\ I=0,\,1; 
$$
$$
\alpha_K^*\approx0.3+0.9t\ (K^*,\ K^{**},\ K^{***}),\ I=1/2,\,\, etc.
$$
or baryonic
$$
\alpha_N(t)\approx-0.3+0.9t\ (N(939),\ N(1688),\ N(2220),...);\ 
$$
$$
\alpha_{\Delta}(t)\approx0.9t,\ (\Delta(1232),\ \Delta(1950),\ \Delta(2400),...),\ etc.  
$$
The
above are simple examples. We use more advanced non-linear trajectories and much higher precision.  

The above trajectories are non-leading, also called ``secondary'', meaning that the relevant cross sections decrease with energy due to their intercepts being lower than one, $\alpha(0)<1.$ Apart from the above non-leading trajectories, there are two more, the Pomeron and the Odderon, with vacuum quantum numbers introduced in the theory to explain the rise of the total cross sections (the Pomeron) and create different cross sections involving particles (e.g., protons) and antiparticles (antiprotons). The Pomeron and the Odderon differ through their $C-$parity, positive for the Pomeron and negative for the Odderon. 

The scattering amplitude with a single Regge-pole exchange is 
\begin{equation}
A(s,t)=\beta(\alpha(t))\eta(\alpha(t))s^{\alpha(t)}
\end{equation} 
where $\beta(\alpha(t))=e^{\alpha(t)}$ is the residue and 
\begin{equation}
\eta(t)=\frac{1+\xi e^{-i\pi\alpha(t)}}{\sin\pi\alpha(t)}
\end{equation}
is the signature factor, $\xi=+1\ {\rm or} -1$. For small $|t|$, the signature factor simplifies to 
$\eta(t)\approx e^{-i\pi\alpha(t)/2}.$

In most of the papers on Regge-poles, one assumes that the input is a simple pole on a linear trajectory, subject to a subsequent iteration procedure based on unitarity, also called a Regge-eikonal. 
We instead start from a dipole and non-linear, complex trajectories. The choice of the Regge dipole  is motivated by unitarity. A Regge dipole in the impact parameter representation of the form $e^{\rho^2/R^2(s)}$
scales with the free parameters being fitted to the data. The dipole (derivative of a simple Regge-pole) produces logarithmically rising cross sections at the unit Pomeron intercept.  

\subsection{Duality}

\subsubsection{The Veneziano Model}
The notion ``duality'' has many meanings and applications. We use it as it was introduced  on the basis of finite energy sum rules (FESR) and realized in the Veneziano model in the late 1960s. It implies the equality of the properly summed  contribution from direct-channel resonance to the smooth Regge asymptotic. This observation was due to numerical calculations by R. Dolen, D. Horn and C. Schmid and subsequent  analytic ones by Ademollo, Veneziano and Virasoro, culminating in a simple explicit model of the amplitude in the form of the Euler Beta function by G. Veneziano \cite{Veneziano}.

The model raised much interest, enthusiasm and hopes in the high-energy community, generating a flood of papers on and around the subject. Promising was (and remains) the  generalization of the elastic amplitude to multi-particle reactions. At the same time, the limitations of the Veneziano Beta function were also evident: it is valid only for real and linear trajectories. Thus, infinitely narrow resonances, excluding scattering, result in the production of real, finite-width resonances. While immediate practical applications seemed to be impossible, the interest in duality and dual models continued in more formal directions, namely in the string theory, supersymmetry, etc., away from practical application. 

Searches for applications in physics continued resulting at least in two modifications. One is the logarithmic dual model, and the other one is a class of dual amplitudes with  Mandelstam Analyticity (DAMA) \cite{DAMA}.  Below, we briefly discuss the main ideas behind duality and present the  main features of DAMA.

An elastic scattering amplitude with an infinite number of poles in the $s$ channel can be written in the form  of  the following integral representation: 
\begin{eqnarray}\label{eq:eq8}
A(s,t)=\int_0^1 dxx^{-\alpha(s)-1}f(s,t,x),
\end{eqnarray}
where $f(s,t,x)$ is regular at $x=0$. By expanding $f(s,t,x)$ in a series near $x=0$ and integrating term-by-term, one obtains
\begin{eqnarray}
A(s,t)=\int_0^1 x^{-\alpha(s)-1}\sum_{k=0}^\infty a_k(t,s)x^k=\sum_{k=0}^\infty \frac{a_k(t,s)}{k-\alpha(s)}.
\end{eqnarray}
To cause the equation to become cross symmetric, we set $f(s,t,x)=(1-x)^{-\alpha(t)-1}g(s,t,x),$ where $g(s,t,x)$ is regular at $x=0$ and $x=1$ and is symmetric with respect to $$(x,s)\longleftrightarrow(1-x,t).$$  In this way, instead of Equation~\ref{eq:eq8}, we obtain
\begin{eqnarray}\label{eq:eq10}
A(s,t)=\int_0^1dx x^{-\alpha(s)-1}(1-x)^{-\alpha(t)-1}g(s,t,x).
 \end{eqnarray}
In the simplest case of $g(s,t,x)=1$, expression Equation~\ref{eq:eq10} reduces to the familiar Euler $B$ function, suggested by G. Veneziano as the scattering amplitude:
\begin{eqnarray}
V(s,t)=\int_0^1dx x^{-\alpha(s)-1}(1-x)^{-\alpha(t)-1}=\frac{\Gamma(-\alpha(s))\Gamma(-\alpha(t))}{\Gamma(-\alpha(s)-\alpha(t))}.
\end{eqnarray}
Its properties are as follows:

1. Crossing symmetry: $A(s,t)=A(t,s)$;

2. Meromorphic: i.e., the only singularities are simple poles in $s$ (and in $t$) at 
$\alpha(s)=n, n=0, 1, 2$ (same in $t$). 

3. Pole decomposition: $V(s,t)$ can be expanded in a pole series both in the $s$ or $t$ channel:
\begin{eqnarray}
V(s,t)=\sum_{n=0}^\infty\frac{1}{n-\alpha(s)}\frac{n+\alpha(t)+1}{n!\Gamma(\alpha(t)+1)}=\sum_{n=0}^\infty\frac{1}{n-\alpha(t)}\frac{n+\alpha(s)+1}{n!\Gamma(\alpha(s)+1)}.
\end{eqnarray}

The Veneziano amplitude is valid only in the narrow resonance approximation, with linear trajectories.

\subsubsection{DAMA} 
DA with MA (DAMA) is defined in \cite{DAMA} as
$$D(s,t)=\int_0^1dx x^{-\alpha(s,x)-1}(1-x)^{-\alpha(t,1-x)-1}.$$
In DAMA, trajectories are replaced by homotopies $\alpha(s,x), \ \alpha(t,1-x)$, where

$$\alpha(s,0)=\alpha(s),\ \ \ \alpha(t,0)=\alpha(t)$$  are physical trajectories and

$$\alpha(s,1)=\alpha_0(s)=a+bs,\ \ \alpha(t,1)=\alpha_0(t),\ \ b>0.$$
The Veneziano limit is: $\alpha(s,x)\rightarrow\alpha(s)=\alpha_0(s).$ For more details, see \cite{DAMA}.

\subsection{Super Broad vs. Narrow Resonance Approximation}
Iteration/unitarization procedures are based on two premises: the choices of the input and of the iteration procedures. The procedures that were borrowed form quantum electrodynamics failed in the strong interaction theory. Below, we sketch an alternative \cite{DAMA}.

For simplicity, let us start with a square-root trajectory both in the $s$ and $t$ channels:
\begin{equation}
\alpha(s)=1-\gamma\sqrt{4m^2-s}
\end{equation}
with a corresponding equation for $\alpha(t)$.

\begin{figure}[H]
\includegraphics[width=0.6\linewidth]{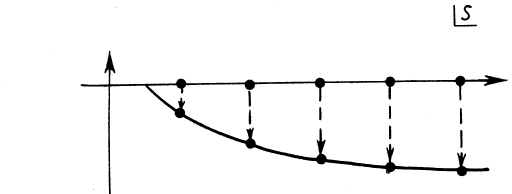}
\caption{Iterating
 the Veneziano amplitude in the complex $s-$plane.}\label{Fig:Veneziano}
\end{figure}

\begin{figure}[H]
\includegraphics[width=0.6\linewidth]{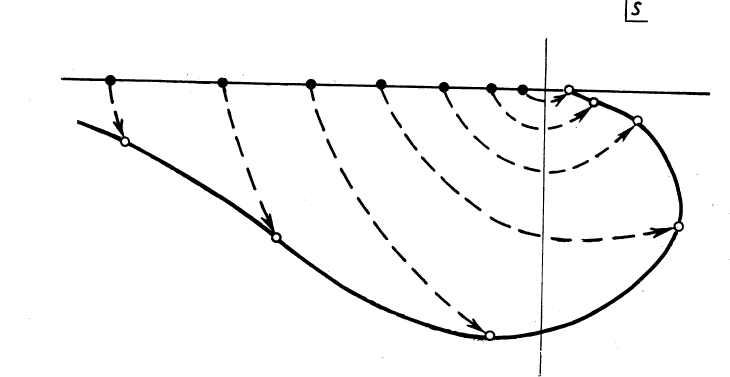}
\caption{Iterating
 DAMA in the complex $s-$plane.}\label{Fig:DAMA}
\end{figure}

While the iteration/unitarization procedure for the Veneziano amplitude aimed to shift the infinite number of poles from the real axis to the non-physical sheet (see Figure~\ref{Fig:Veneziano}), in DAMA, it will move some poles from the real negative axis to the fourth quadrant of the  non-physical $s$ plane, as shown in Figure \ref{Fig:DAMA}, producing a finite number of resonances. 


Note that while in the Veneziano model, it it impossible to shift the poles from the real axis, e.g., by adding a non-linear part to the trajectories without damaging it, this is possible in DAMA, which not only allows but even requires complex trajectories. This procedure is similar to the transition from a Born series in coupling constant $g$ to the ``anti-Born'' expansion in the inverse coupling  $1/g$.

Duality in our approach is essential in the inelastic structure functions, where the dependence on the missing mass via the direct-channel trajectory $\alpha(M^2)$ connects the low-mass, resonance region (resonances) with its Regge asymptotic.

\section{Regge Dipole Model of Elastic Scattering}\label{sec:dipole_pom_odd}

The basic object of this approach is a double Regge--Pomeron pole, alternatively called a dipole Pomeron (DP) \cite{Bialkowski, Phillips, JVall}. This object is the unique alternative to a simple pole. It has a number of advantages over the simple pole; in particular, it produces logarithmically rising cross sections at an intercept of one. It is also unique since higher-order poles are forbidden by unitarity, since they would violate unitarity. Furthermore, the dipole Pomeron scales and is invariant against $s-$channel unitarization, leaving its form invariant, merely rescaling the parameters that are already adjusted to the data. 
These are the main properties of the DP used as~input. 

In early 1970s, a dip-bump structure was observed in proton--proton scattering at the ISR \cite{Nagy}. A single dip followed by a maximum (bump) was observed at other proton accelerators. The phenomenon was a puzzle since it was not clear whether it results from the composite structure of the nucleon (rescattering of quarks) or from geometrical considerations (shape of the nucleon) or just from unitarity, as calculated, e.g., by means of eikonal corrections, known also as a Regge eikonal \cite{Peter}. While most of these approaches predicted many dips and bumps, experimentally, a single minimum (dip) followed by maximum (bump) was observed. At the new energy frontier, the LHC confirmed this trend, leaving the puzzle unresolved.  

A single Regge--Pomeron pole model does not produce any structure; rather, it produces a smooth diffraction cone, the exponential shape being slightly modified by non-linear trajectories or $t-$channel unitarity corrections.  In Ref. \cite{JVall}, the DP model was appended by a dip-bump mechanism based on a non-trivial interference between the simple and double poles moving with energy. Remarkably, the extended DP model does not generate extra dips and bumps, and it fits the data. 

     In view of the above considerations, we find it important to investigate the existence of similar dip-bump structures in proton diffraction dissociation. We start with the better known single diffraction dissociation (SD). This reaction was studied in detail in Ref.~\cite{Kuprash} and reviewed in \cite{Barone,Kaidalov}. We start by reviewing the dip-bump mechanism in the DP model applied to elastic scattering to be extended to SD in the subsequent section.   
    
The dipole Pomeron (DP) amplitude extended by the dip-bump mechanism, mentioned above,  is defined as \cite{JVall}\vspace{-3pt}

\begin{align}\label{Pomeron}
	A_P(s,t)&={d\over{d\alpha_P}}\Bigl[{\rm e}^{-i\pi\alpha_P/2}G(\alpha_P)\Bigl(s/s_{0P}\Bigr)^{\alpha_P}\Bigr]\\\nonumber &= 
	{\rm e}^{-i\pi\alpha_P(t)/2}\Bigl(s/s_{0P}\Bigr)^{\alpha_P(t)}\Bigl[G'(\alpha_P)+\Bigl(L_P-i\pi
	/2\Bigr)G(\alpha_P)\Bigr],
	\end{align}
where $L_P=\ln{\left(s/s_{0P}\right)}$, and $s$ and $t$ are the standard Mandelstam variables. Since $G'$ in the second squared brackets determines the shape of the cone, we fix
	\begin{equation} \label{residue} G'(\alpha_P)=-a_P{\rm
		e}^{b_P[\alpha_P-\alpha_{0P}]},\end{equation}
  where $\alpha_{0P}$ is the intercept of $\alpha_P$.  $G(\alpha_P)$ is recovered
	via integration:
	\begin{equation} \label{residue_int} G(\alpha_P)=\int d \alpha_P G'(\alpha_P) =-a_P\left({\rm
		e}^{b_P[\alpha_P-\alpha_{0P}]}/b_P-\gamma_P\right).\end{equation}
By introducing the parameter $\epsilon_P=\gamma_Pb_P$, the Pomeron amplitude Equation~(\ref{Pomeron}) can be rewritten in a ``geometrical`` form:
	\begin{equation}\label{GP}
	A_P(s,t)=i\frac{a_P}{b_P}\left(\frac{s}{s_{0P}}\right)^{\alpha_{0P}}e^{-\frac{i\pi}{2}\left(\alpha_{0P}-1\right)}\left[r_{1P}^2{\rm e}^{r^2_{1P}\left[\alpha_P(t)-\alpha_{0P}\right]}-\varepsilon_P r_{2P}^2{\rm e}^{r^2_{2P}\left[\alpha_P(t)-\alpha_{0P}\right]}\right],
	\end{equation} 
	where $r_{1P}^2(s)=b_P+L_P-i\pi/2$ and $r_{2P}^2(s)=L_P-i\pi/2$. 
The Odderon contribution is
\begin{eqnarray}\label{eq:DP_Odderon}
A_O(s,t)=-iA_{P\rightarrow O}(s,t).
\end{eqnarray} 


We use the norm where
\begin{equation}
\sigma_{\rm tot}(s)=\frac{4\pi}{s}\, {\rm Im} \, A(s,t=0),
\label{eq:total_cross_section}
\end{equation}
\begin{equation}
\frac{{\rm d}\sigma_{\rm el}}{{\rm d} t}(s,t)=\frac{\pi}{s^2}\left|A\left(s,t\right)\right|^2.
\label{eq:differential_cross_section}
\end{equation}

\begin{figure}[hbt!]
\includegraphics[clip,scale=0.6]{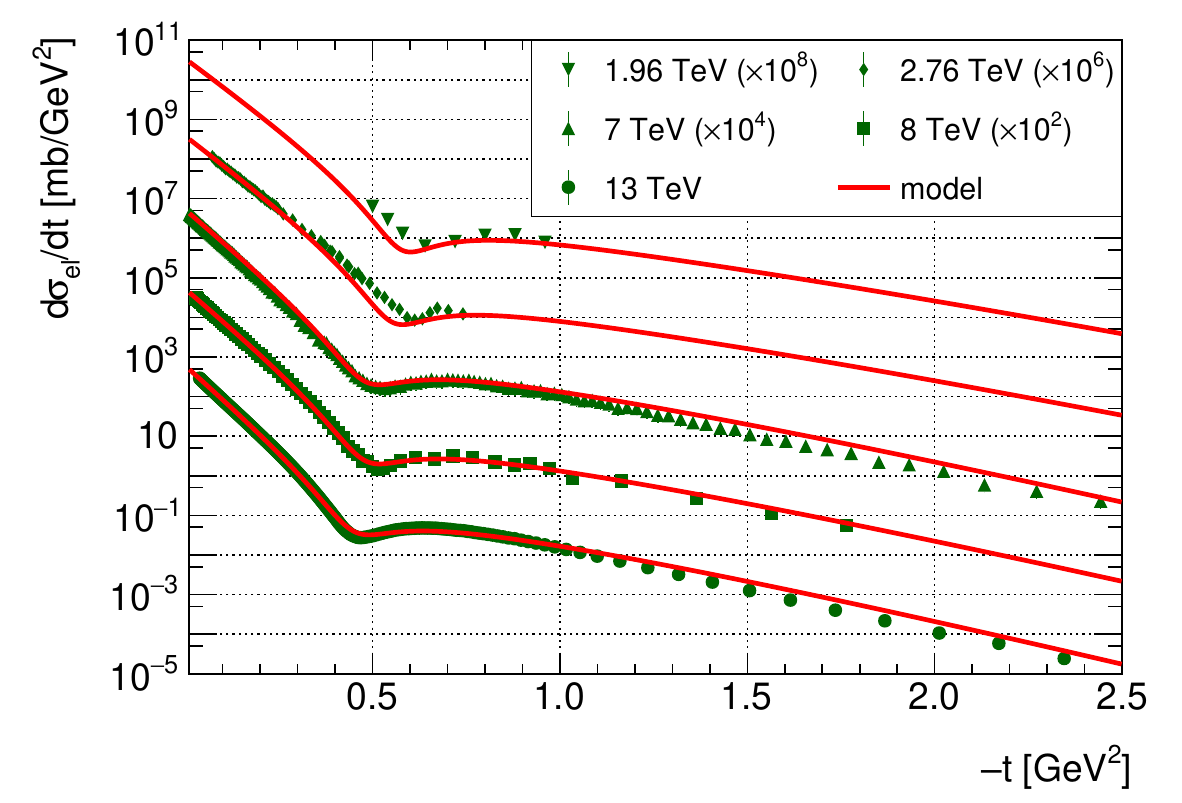}
\caption{Differential
 cross section of elastic $pp$ scattering at TeV energies in the dipole Pomeron and Odderon model.}
 \label{fig:dsig_DP_pp}
\end{figure}

\begin{figure}[hbt!]
\includegraphics[clip,scale=0.6]{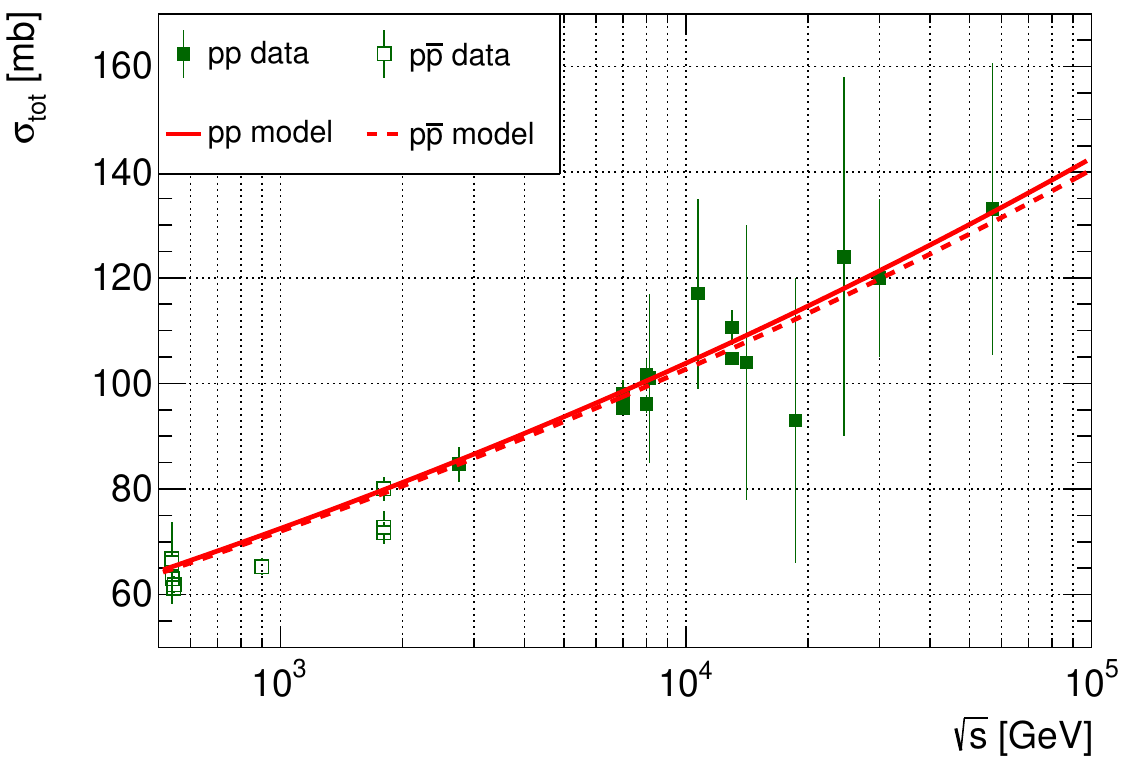}
\caption{Total
 cross section of $pp$ and $p\bar p$ scattering at TeV energies.}
\label{fig:sig_tot_DP}
\end{figure}

\begin{figure}[hbt!]
\includegraphics[clip,scale=0.6]{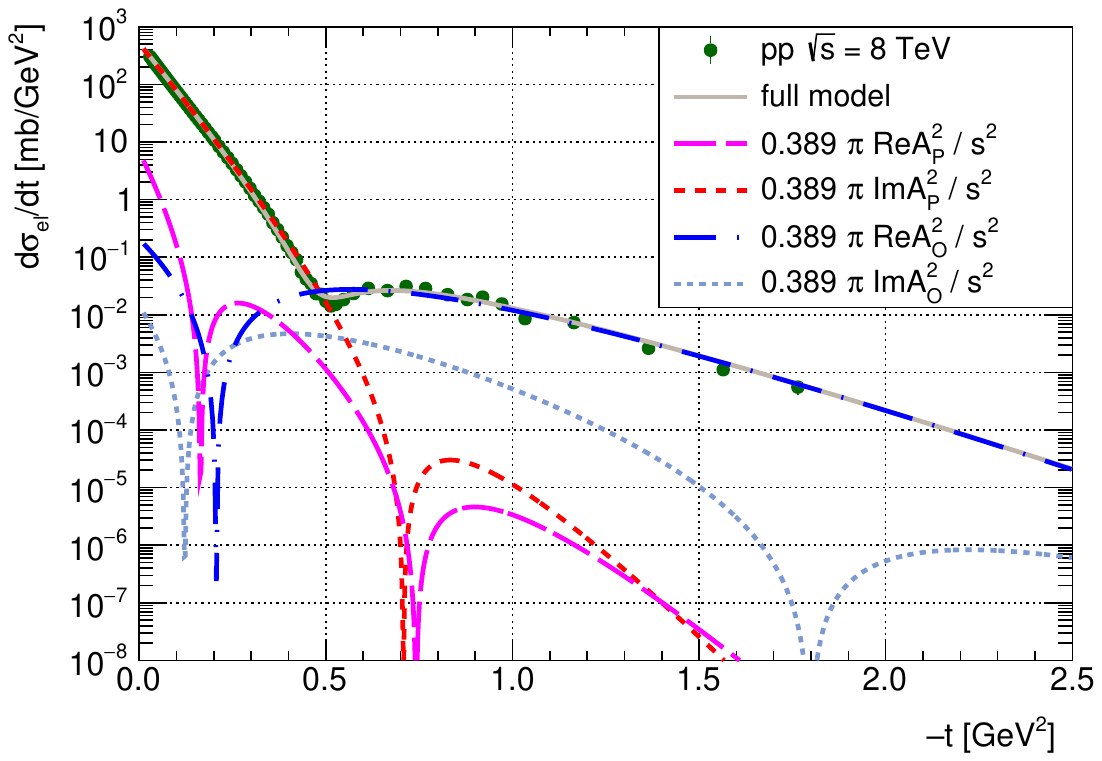}
\caption{Contribution
 of various components of the amplitude to the differential cross section at $\sqrt{s}=8$ TeV in the dipole Pomeron (DP) and Odderon models.}
\label{fig:dsig_DP_comps}
\end{figure}

Typical fits to the data by the Regge dipole Pomeron and Odderon model are shown in Figure~\ref{fig:dsig_DP_pp}  and Figure~\ref{fig:sig_tot_DP}.
The values of the fitted parameters are given in Table~\ref{tab:DP_PO_fit_pars}. Figure~\ref{fig:dsig_DP_comps} shows the contribution of particular components in the differential cross section at \mbox{$\sqrt{s}=8$~TeV.} 

\begin{table}[hbt!]    
	\caption{Parameters fitted to $pp$ and $p\bar p$ data on elastic differential cross section, total cross section and the ratio~$\rho$. }
	\setlength{\tabcolsep}{9.6mm}
 \resizebox{\linewidth}{!}{\begin{tabular}{cccc}
		\toprule
	\multicolumn{2}{c}{\textbf{Pomeron}}&\multicolumn{2}{c}{\textbf{Odderon}} \\
		\midrule
		$\delta_P$ & $0.02865$ &$\delta_O$&$0.2042$  \\
		$\alpha^{'}_{P}~[{\rm GeV}^{-2}]$ & $0.4284$  &$\alpha^{'}_{O}~[{\rm GeV}^{-2}]$ &$0.1494$  \\
		$a_P$ & $45.63$ &$a_O$&$0.01934$ \\
		$b_P$ & $4.873$  &$b_O$ &$2.160$         \\
		$\gamma_P$ & $0.06085$  &$\gamma_O$&$0.4866$  \\
		$s_{0P}~[{\rm GeV}^2]$ & $11.26$ &$s_{0O}~[{\rm GeV}^2]$ & $1.026$ \\ 
	\bottomrule
	\end{tabular}}
	\label{tab:DP_PO_fit_pars}
\end{table}\vspace{-9pt}

\vspace{-9pt}

\section{Dip and Bump in Elastic Scattering}\label{sec:DP_min_max}

\begin{figure}[H]
\includegraphics[width=0.7\textwidth]{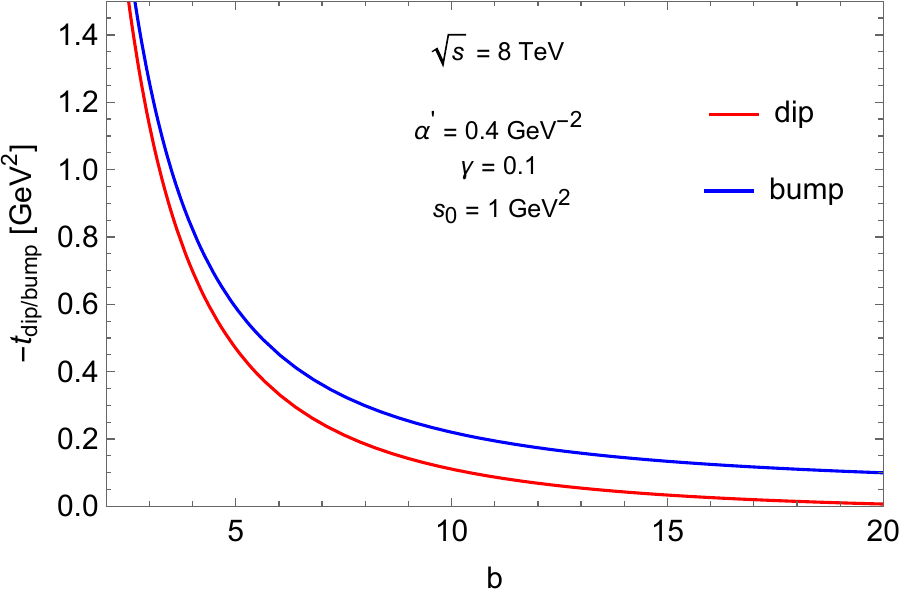}
\caption{Position
 of the dip and bump as function of $b$.}
\label{fig:b_dep_of_db}
\end{figure}

A dipole amplitude generates a dip-bump structure in the differential cross section.  The positions of the minimum (dip) and the maximum (bump) of the elastic differential cross section are
\begin{equation}
    -t_{dip} =\frac{1}{\alpha^{'}b}\ln \frac{(b+L)}{\gamma b L}
\end{equation}
and
\begin{equation}
    -t_{bump} =\frac{1}{\alpha^{'}b}\ln \frac{\left[4(b+L)^2+\pi^2\right]}{\gamma b (4L^2+\pi^2)}
\end{equation}
Figure~\ref{fig:b_dep_of_db} shows the position of the dip and the bump as a function of the  slope $b$. This parameter is correlated with the slope of the differential cross section. The smaller the value of the $b$ parameter, the higher the $|t|$ value where the dip-bump structure appears. The slope of the $-t$ distribution decreases by about a factor of two when we go from elastic scattering to single diffractive dissociation. Thus, in the picture provided by the dipole Regge framework, it is natural to expect that in single diffractive dissociation, a possible dip-bump structure appears at higher $|t|$ values than it does in elastic scattering. 

It is known that in $pp$ elastic scattering, the slope of the diffraction cone is energy dependent and it rises with increasing energy. As the energy rises, the position of the dip-bump structure moves to smaller $-t$ values. The energy-dependent slope is given by the derivative of the logarithm of the differential cross section at $t=0$.  In the case of single diffraction dissociation, the slope of the differential cross section depends not only on the energy but also on the mass squared of the produced hadronic system, as detailed below.


Note that the above dip-bump mechanism does not depend on the form of the exchanged trajectory. For small $|t|$, a linear trajectory can be used, while in the missing mass channel, the non-linearity of the complex trajectory 
$\alpha(M^2)$ is essential in providing finite widths of resonances.

\section{Single Diffractive Dissociation (SD)}\label{sec:SD_dipole}

Proton's diffractive dissociations, both simple (SD) and double (DD), were studied in many papers \cite{Kaidalov, Goulianos, Collins, JL,Donnachie:1984xq, triple}. In a recent paper \cite{Martin}, SD was considered within the Good--Walker formalism. Calculations of a particular case 
of central $\rho^0$ production in pp collisions with single proton diffractive dissociation at the LHC can be found in a recent paper \cite{Szczur}. To our best knowledge, possible structures in the $t$ distribution of SD, DD or CED have not been studied until now.  

To generate a dip-bump structure in single diffraction dissociation, we introduce a dipole Pomeron (DP) and Odderon exchange to the differential cross section. 

In the triple Regge approach (see \cite{Kaidalov, Goulianos, Collins, JL,Donnachie:1984xq, triple}), the triple PPP  exchange contribution to the SD differential cross section is
\begin{align}
\frac{d^2\sigma_{SD}}{dtdM^2}&=\frac{1}{16\pi^2}\frac{1}{M^2}g_{Ppp}^2(t)\left(s/M^2\right)^{2\alpha_P(t)-2}g_{PPP}(t)g_{Ppp}(0)\left(M^2\right)^{\delta_P}.
\end{align}
$g_{PPP}$ is nearly $t$-independent  \cite{Barone}.  Thus, for the $t$-dependent part of the amplitude of SD, we have
\begin{equation}
  A^{SP}_{SD}(s,M^2,\alpha_P)  \sim \eta(\alpha_P)G_P(\alpha_P)\left(s/M^2\right)^{\alpha_P}
\end{equation}
where the $t$-dependence resulting from $g_{Ppp}(t)$ is accounted for by $G(\alpha)$. Hence  the $t$-dependent part of the dipole Pomeron amplitude is
\begin{align}
A^{DP}_{SD}(s,M^2,\alpha_P)&={d\over{d\alpha_P}}A^{SP}_{SD}(s,M^2,\alpha)  \\ \nonumber
&\sim{\rm e}^{-i\pi\alpha/2}\left(s/M^2\right)^{\alpha}\left[G_P'(\alpha_P)+\Bigl(L_{SD}-i\pi
/2\Bigr)G_P(\alpha_P)\right],
\end{align}
where
\begin{equation}
    L_{SD} \equiv \ln{\left(s/M^2\right)}.
\end{equation}
The resulting double differential cross section for the SD process is
\begin{align}\label{eq:SDdiff_PPP}
    \frac{d^2\sigma_{SD}^{PPP}}{dtdM^2}&=\frac{1}{M^2}\left(G_P'^2(\alpha_P)+2L_{SD}G_P(\alpha_P)G_P'(\alpha_P) + G_P^2(\alpha_P)\left(L^2_{SD}+\frac{\pi^2}{4}\right)\right)\times \\\nonumber&\times\left(s/M^2\right)^{2\alpha_P(t)-2}\sigma^{Pp}(M^2),
\end{align}
where
\begin{equation}\label{eq:Pptot}
  \sigma^{Pp}(M^2) = g_{PPP}g_{Ppp}(0)\left(M^2\right)^{\alpha(0)-1}.
\end{equation}
By using the  proton relative momentum loss variable $\xi=M^2/s$, we obtain
\begin{equation}
    \frac{d^2\sigma_{SD}^{PPP}}{dtd\xi}=\left(G'^2(\alpha)+2L_{SD}G(\alpha)G'(\alpha) + G^2(\alpha)\left(L^2_{SD}+\frac{\pi^2}{4}\right)\right)\xi^{1-2\alpha(t)}\sigma^{Pp}(s\xi)
\end{equation}
where
\begin{equation}
    L_{SD} \equiv  -\ln\xi
\end{equation}

Using Equations~(\ref{residue}) and   (\ref{residue_int}),    
one finds the position of the dip and the bump in the SD differential cross section:
\begin{equation}\label{eq:dip_pos}
    t^{SD}_{dip} = \frac{1}{\alpha'b}\ln\frac{\gamma b L_{SD}}{b+L},
\end{equation}
\begin{equation}\label{eq:bump_pos}
    t^{SD}_{bump} = \frac{1}{\alpha'b}\ln\frac{\gamma b( 4L_{SD}^2+\pi^2)}{4(b+L_{SD})^2+\pi^2}.
\end{equation}

We also include the contribution of the Odderon in form of the Odderon--Odderon--Pomeron (OOP) triple exchange. In standard triple Regge formalism, it takes the following form:
\begin{align}
\frac{d^2\sigma_{SD}^{OOP}}{dtdM^2}&=\frac{1}{16\pi^2}\frac{1}{M^2}g_{Opp}^2(t)\left(s/M^2\right)^{2\alpha_O(t)-2}g_{OOP}(t)g_{Ppp}(0)\left(M^2\right)^{\delta_O}
\end{align}
We   use its dipole form (the derivation is the same as in case of the dipole PPP contribution):
\begin{align}\label{eq:SDdiff_OOP}
    \frac{d^2\sigma_{SD}^{OOP}}{dtdM^2}&=\frac{1}{M^2}\left(G_O'^2(\alpha_O)+2L_{SD}G_O(\alpha_O)G_O'(\alpha_O) + G_O^2(\alpha_O)\left(L^2_{SD}+\frac{\pi^2}{4}\right)\right)\times\\\nonumber&\times\left(s/M^2\right)^{2\alpha_O(t)-2}\sigma^{Pp}(M^2),
\end{align}
where $\sigma^{Pp}$ is given by Equation~(\ref{eq:Pptot}), and the possible difference between $g_{OOP}$ and $g_{PPP}$ is accounted for by  $a_O$. The dip and bump positions in the dipole OOP contribution are given by Equations~(\ref{eq:dip_pos}) and~(\ref{eq:bump_pos}).

The complete model for double differential cross section of SD is a sum of three triple Regge contributions appended by a pion exchange contribution: 
\begin{equation}\label{eq:SDdiff_full}
    \frac{d^2\sigma_{SD}}{dtdM^2}=\frac{d^2\sigma_{SD}^{PPP}}{dtdM^2}+\frac{d^2\sigma_{SD}^{OOP}}{dtdM^2}+\frac{d^2\sigma_{SD}^{RRP}}{dtdM^2}+\frac{d^2\sigma_{SD}^{\pi}}{dtdM^2}
\end{equation}

The values of the parameters are summarized in Table~\ref{tab:SD_DP_PO_fit_pars}. The parameters $a$ and $b$ were fitted to the data. The  parameter $\gamma$ is fixed at the value obtained in the analysis of the elastic scattering data. The intercepts of the Pomeron and the Odderon trajectories are 1, i.e., $\delta_P=\delta_O=0$. The slopes of the Pomeron and the Odderon trajectories are fixed at the values obtained in the analysis of the elastic scattering data.

 Interestingly, the DP Pomeron model of proton--proton SD produces properly rising total integrated cross sections with a unit Pomeron intercept, i.e., with $\delta=0$. 
 
\begin{table}[H]    
	\caption{Values of the parameters of the model for SD cross sections.}
	\setlength{\tabcolsep}{3.3mm}
 \resizebox{\linewidth}{!}{\begin{tabular}{cccccccc}
		\toprule
		\multicolumn{1}{c}{}&\multicolumn{2}{c}{\textbf{Pomeron}}&\multicolumn{2}{c}{\textbf{Odderon}}&\multicolumn{2}{c}{\textbf{Reggeon}}&\multicolumn{1}{c}{} \\
		\midrule
		&$\delta_P$ & $0$ &$\delta_O$&$0$ &$\delta_R$ &$-0.4500$ & \\
		&$\alpha^{'}_{P}~[{\rm GeV}^{-2}]$ & $0.4284$  &$\alpha^{'}_{O}~[{\rm GeV}^{-2}]$ &$0.1494$&$\alpha^{'}_{R}~[{\rm GeV}^{-2}]$ &$0.9300$ & \\
		&$a_P$ & $45.63$ &$a_O$&$0.01934$&$a_R$&$2.500$& \\
		&$b_P$ & $4.873$  &$b_O$ &$2.160$&$b_R$&$0$ &    \\
		&$\gamma_P$ & $0.06085$  &$\gamma_O$&$0.4866$ &$-$&$-$&\\
		\bottomrule
	\end{tabular}}
	\label{tab:SD_DP_PO_fit_pars}
\end{table}

\section{Dip-Bump in SD}\label{sec:SD_db}

In this section, we show our predictions concerning the dip-bump structure in single diffractive dissociation at SPS and LHC.  

\begin{figure}[H]
\includegraphics[width=.63\textwidth]{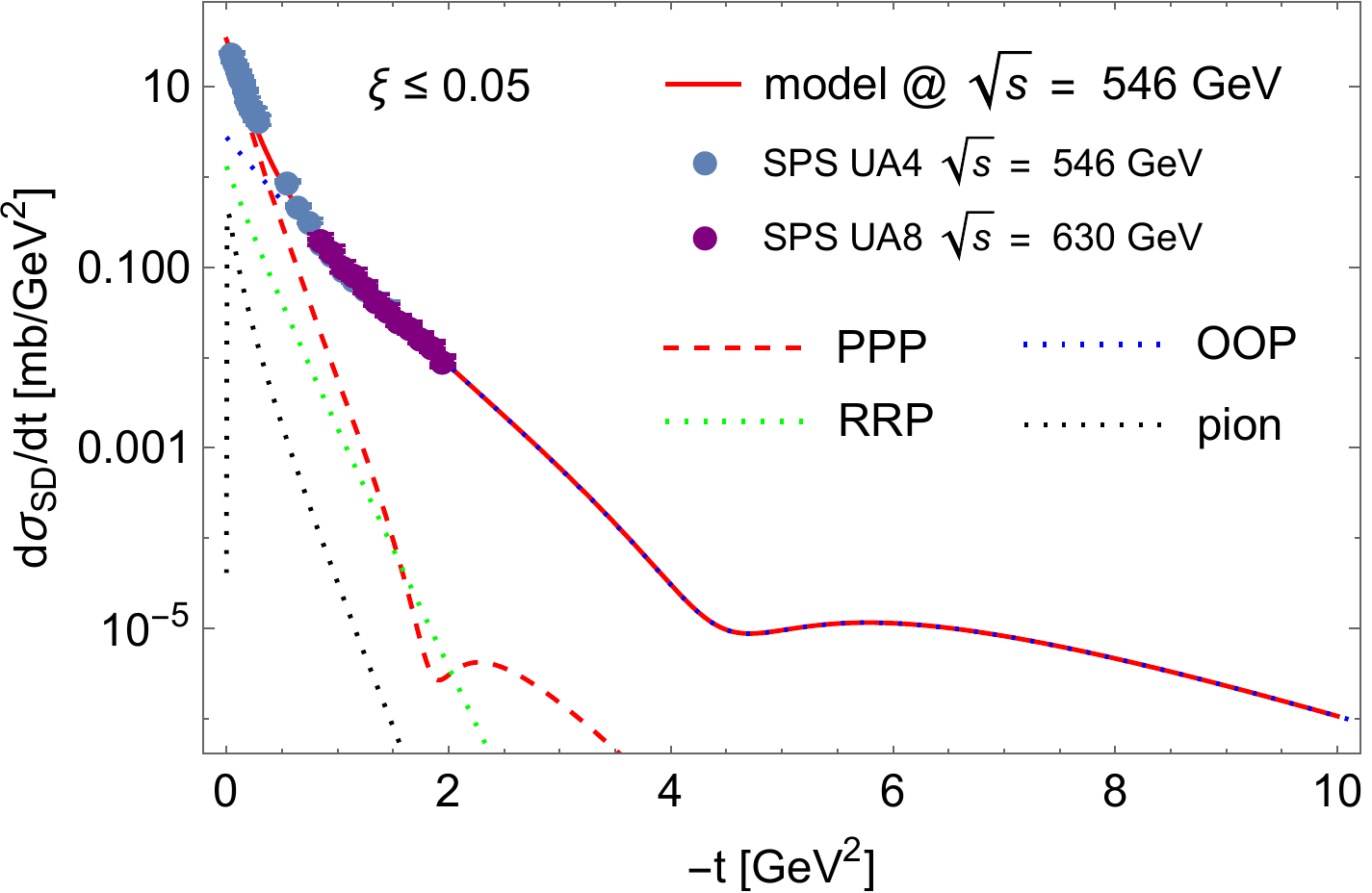}
\caption{Predicted dip-bump structure in the $t$ distribution of the $\xi$ integrated differential cross section of SD at $\sqrt{s}=$ 546 GeV, with various contributions shown separately.}
\label{fig:SD_ds_SPS_xi_int_t_dep_db}
\end{figure}

The predicted structure in the $t$ distribution of the differential cross section at \linebreak  \mbox{$\sqrt{s}=$ 546 GeV}  integrated in $\xi$ is shown in Figure~\ref{fig:SD_ds_SPS_xi_int_t_dep_db}. 

\begin{figure}[H]
\includegraphics[width=0.63\textwidth]{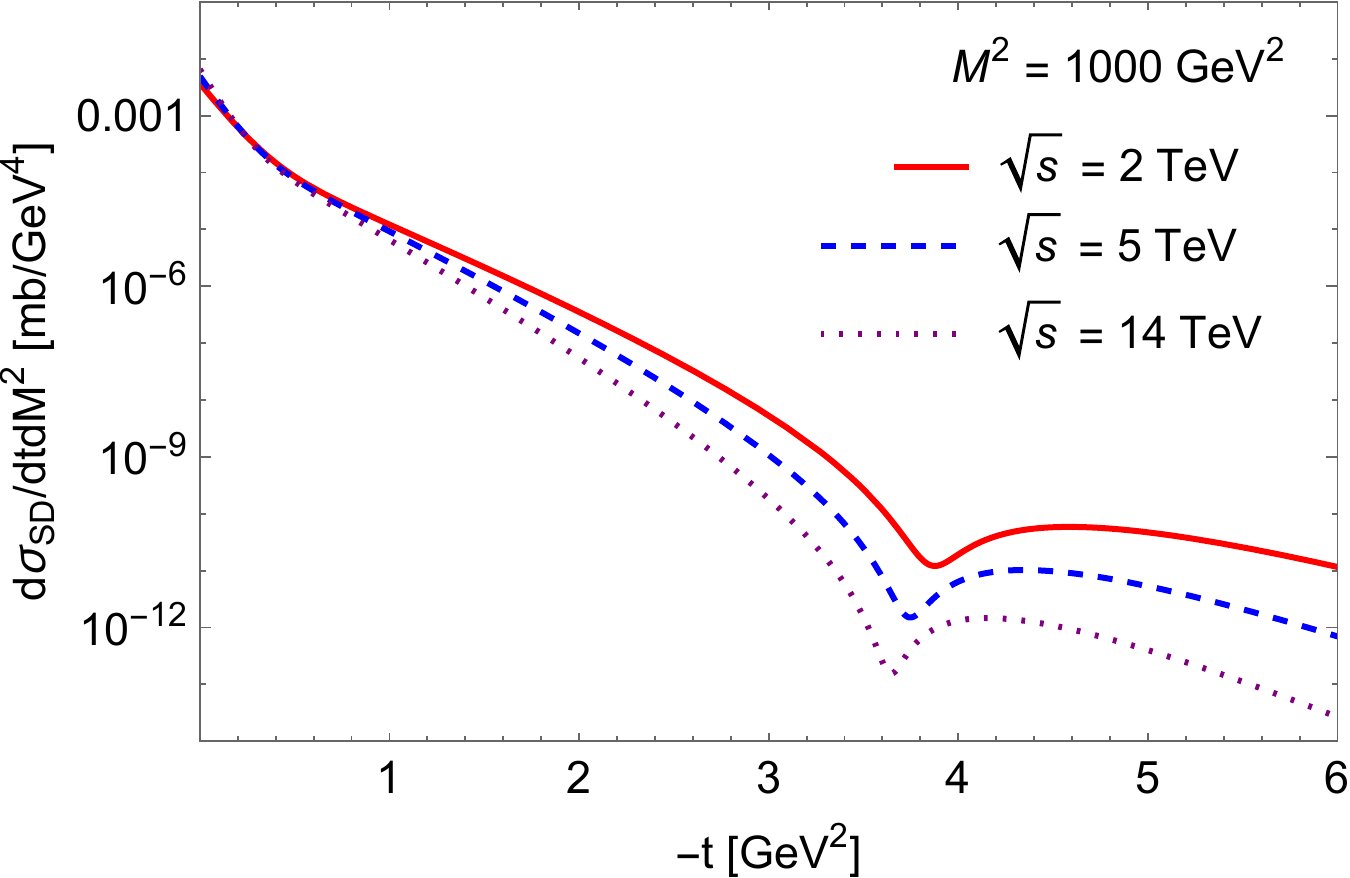}
\caption{Prediction of the energy dependence of the dip-bump structure in the $t$
distribution of the double differential cross section of SD at $M^2=$ 1000 GeV$^2$.}
\label{fig:SD_ds_s_dep_M2_fix_t_dep}
\end{figure}

Figure~\ref{fig:SD_ds_s_dep_M2_fix_t_dep} shows the dip-bump structure at fixed $M^2$ value. One can see that with increasing energy and fixed $M^2$ value or with decreasing $M^2$ at a fixed energy, the position of the dip-bump structure goes to smaller $-t$ values. The latter property is related to the $M^2$-dependence of the slope of the $t$ distribution of the differential cross section of the single diffractive dissociation. As $M^2$ increases, the slope of the  $t$ distribution decreases and the dip-bump structure moves to higher $|t|$ values (see below).

Figure~\ref{fig:SD_ds_630_GeV_xi_fix_t_dep_db} shows the prediction at $\xi=0.05$ and $\sqrt{s}=$ 630 GeV. 
 Different contributions to the differential cross section are also shown. One can see that both the dipole PPP triple exchange and the dipole OOP triple exchange generate a dip-bump structure. The experimentally observable effect is expected from the OOP contribution.

\begin{figure}[H]
\includegraphics[width=0.63\textwidth]{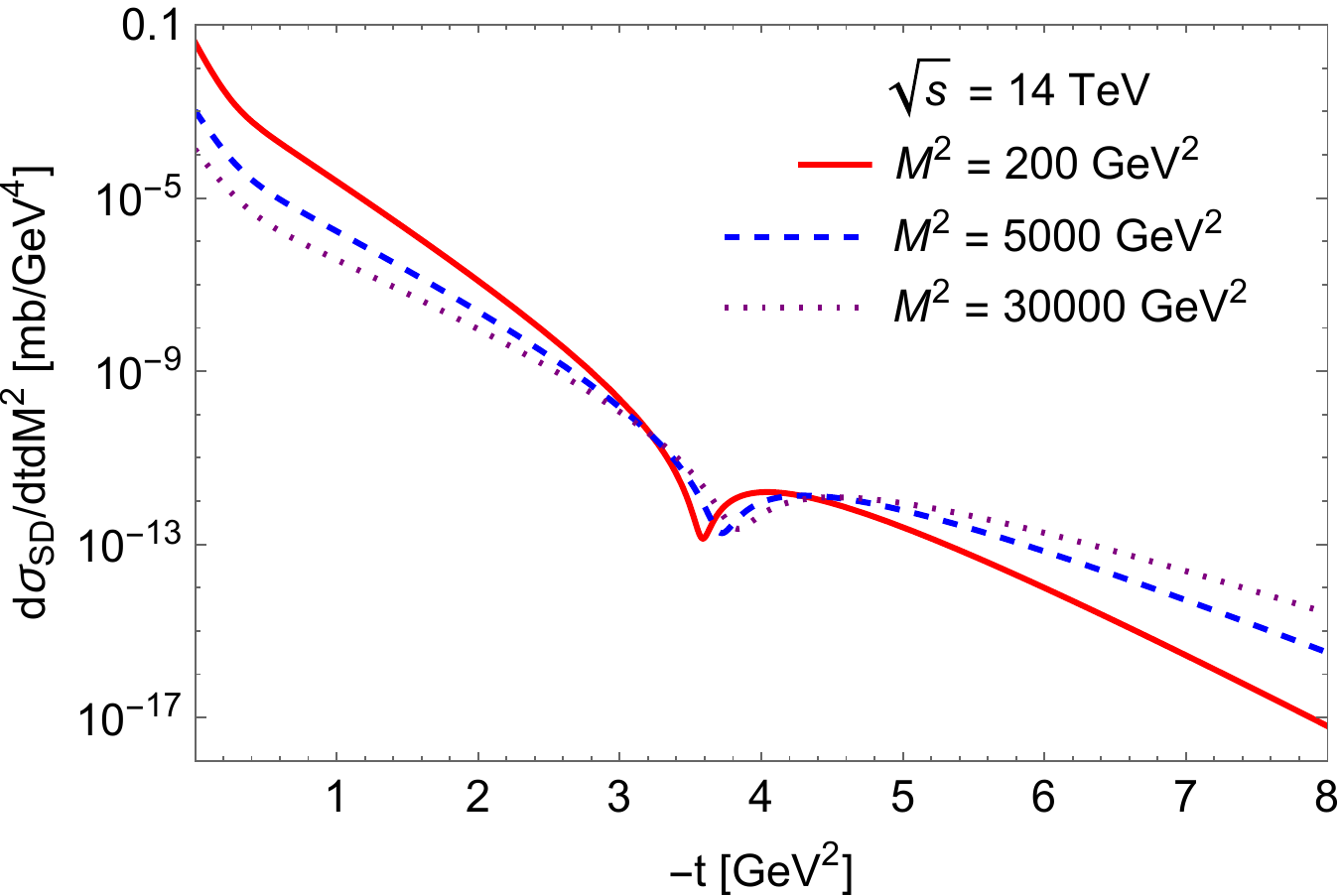}
\caption{Our
 prediction of the $M^2$ dependence of the dip-bump structure of the $t$ distribution of the double differential cross section of SD at $\sqrt{s}=$ 14 TeV.}
\label{fig:SD_ds_14_TeV_M2_fix_t_dep}
\end{figure}

The predicted dip-bump at $\sqrt s = 14$ TeV for selected missing masses $M^2$ is shown in Figure~\ref{fig:SD_ds_14_TeV_M2_fix_t_dep}. 
\begin{figure}[H]
\includegraphics[width=0.63\textwidth]{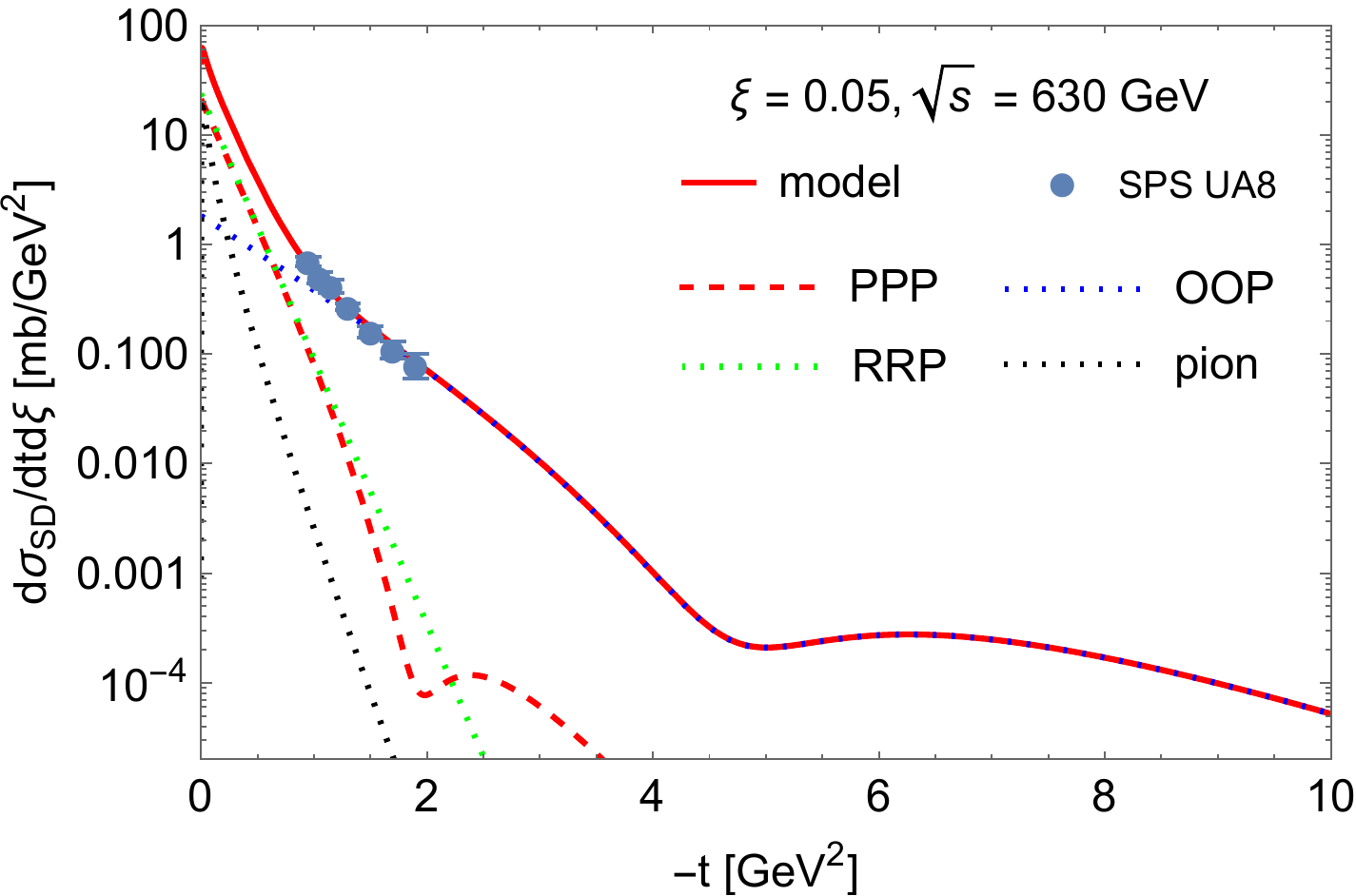}
\caption{Predicted
 dip-bump structure in the $t$ distribution of the double differential cross section of SD  at $\xi=0.05$ and $\sqrt{s}=$ 630 GeV. }
\label{fig:SD_ds_630_GeV_xi_fix_t_dep_db}
\end{figure}\vspace{-9pt}

\begin{figure}[H]
\includegraphics[width=0.63\textwidth]{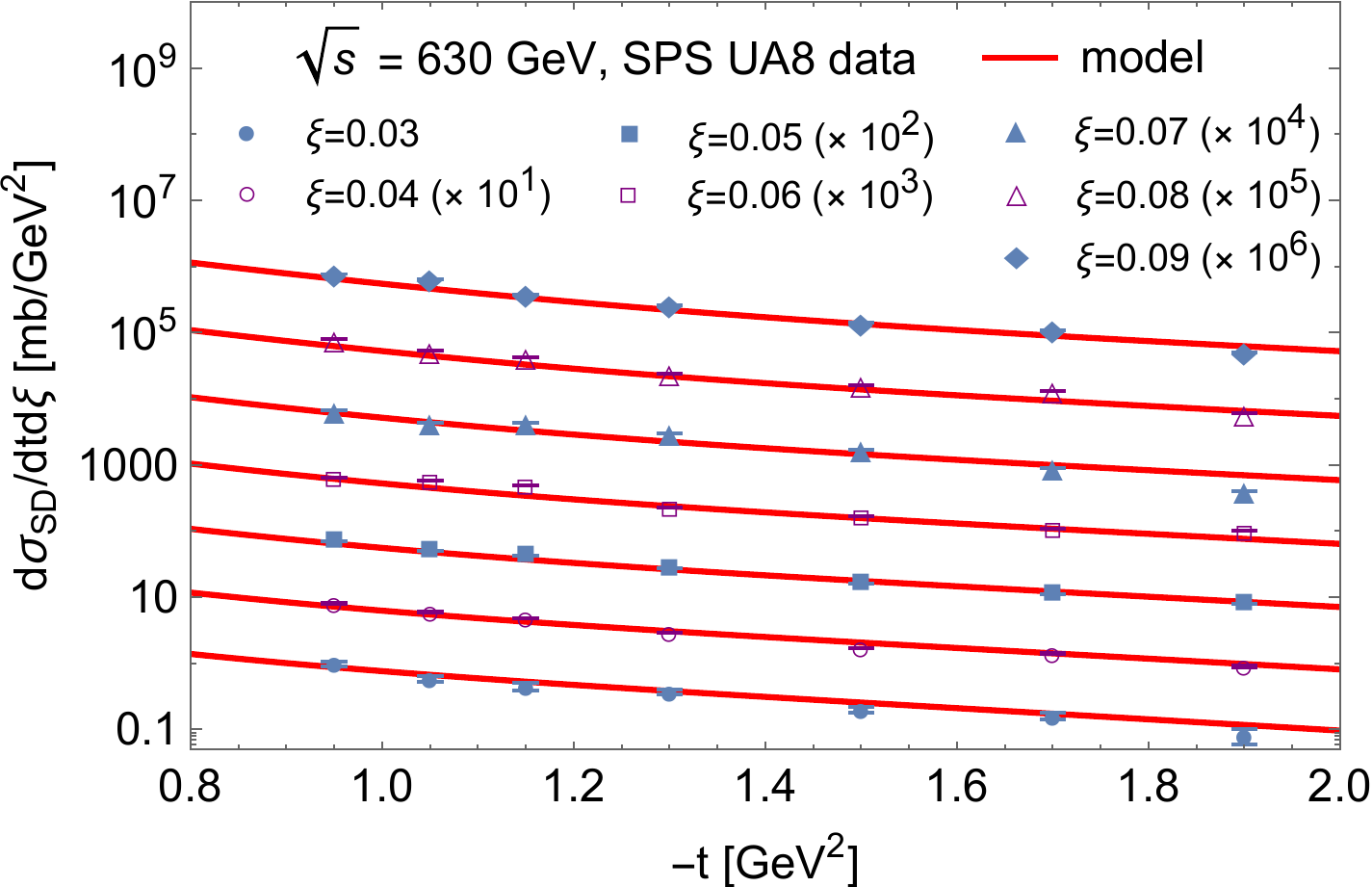}
\caption{Predicted double differential cross section of proton--antiproton SD at fixed values of $\xi$ and at $\sqrt{s}=630$ GeV as function of $-t$.}
\label{fig:SD_ds_SPS_GeV_fix_xi_t_dep}
\end{figure}

Our descriptions for double differential cross section of proton--antiproton SD  at fixed $\xi$ and $\sqrt{s}=630$ GeV as a function of $-t$  are shown in Figure 
\ref{fig:SD_ds_SPS_GeV_fix_xi_t_dep}.

\begin{figure}[H]
\includegraphics[width=0.63\textwidth]{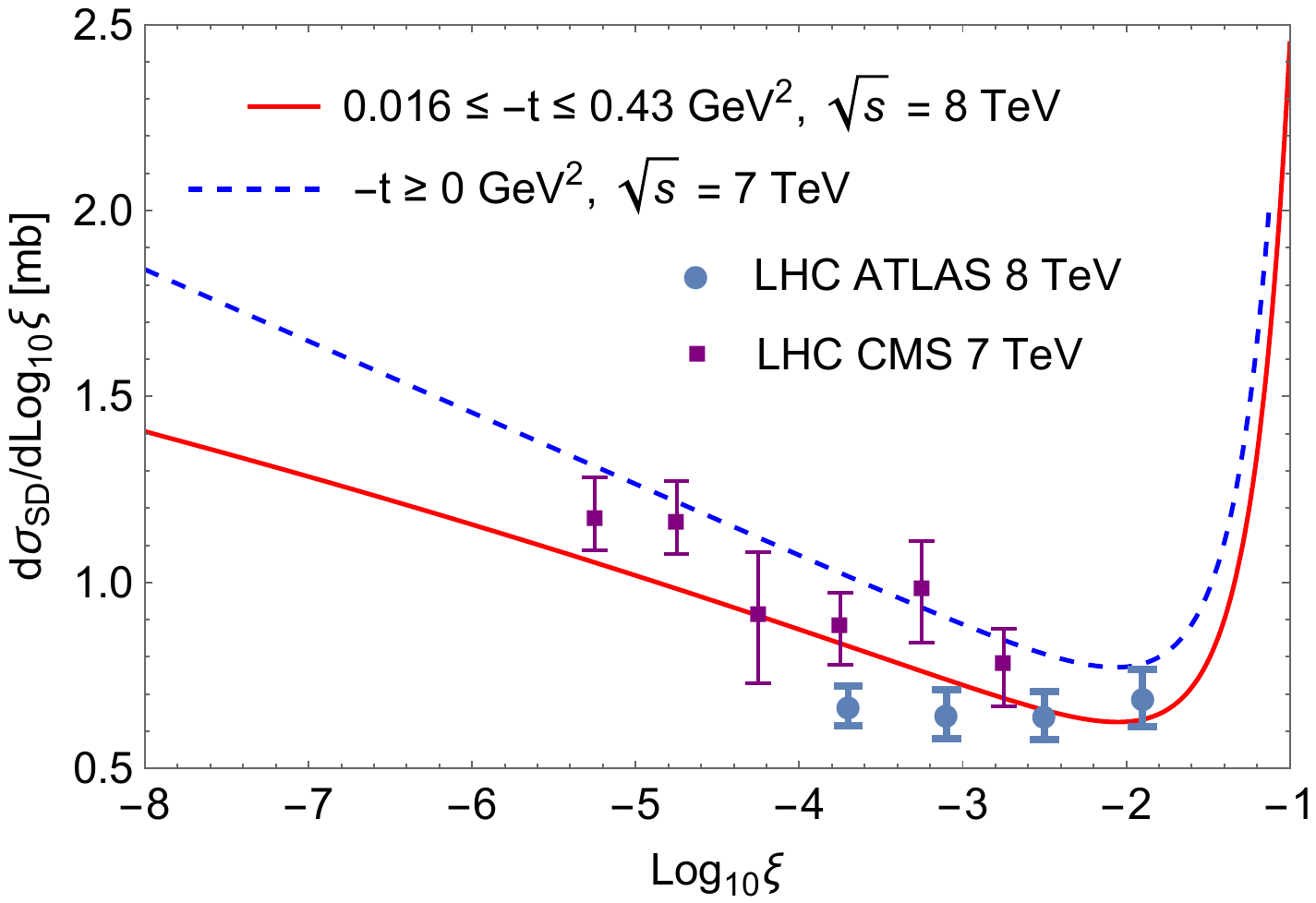}
\caption{Integrated differential cross section of  proton--proton SD  at $\sqrt{s}=7\ \rm{and}\ 8$ TeV.}
\label{fig:SD_ds_ATLAS_CMS_t_int_xi_dep}
\end{figure}

Predictions for the integrated differential cross section of  proton--proton SD at $\sqrt{s}=8$ TeV as a function of $-t$ are shown in Figure 
\ref{fig:SD_ds_ATLAS_CMS_t_int_xi_dep}.

\vspace{-9pt}

\begin{figure}[H]
\includegraphics[width=0.63\textwidth]{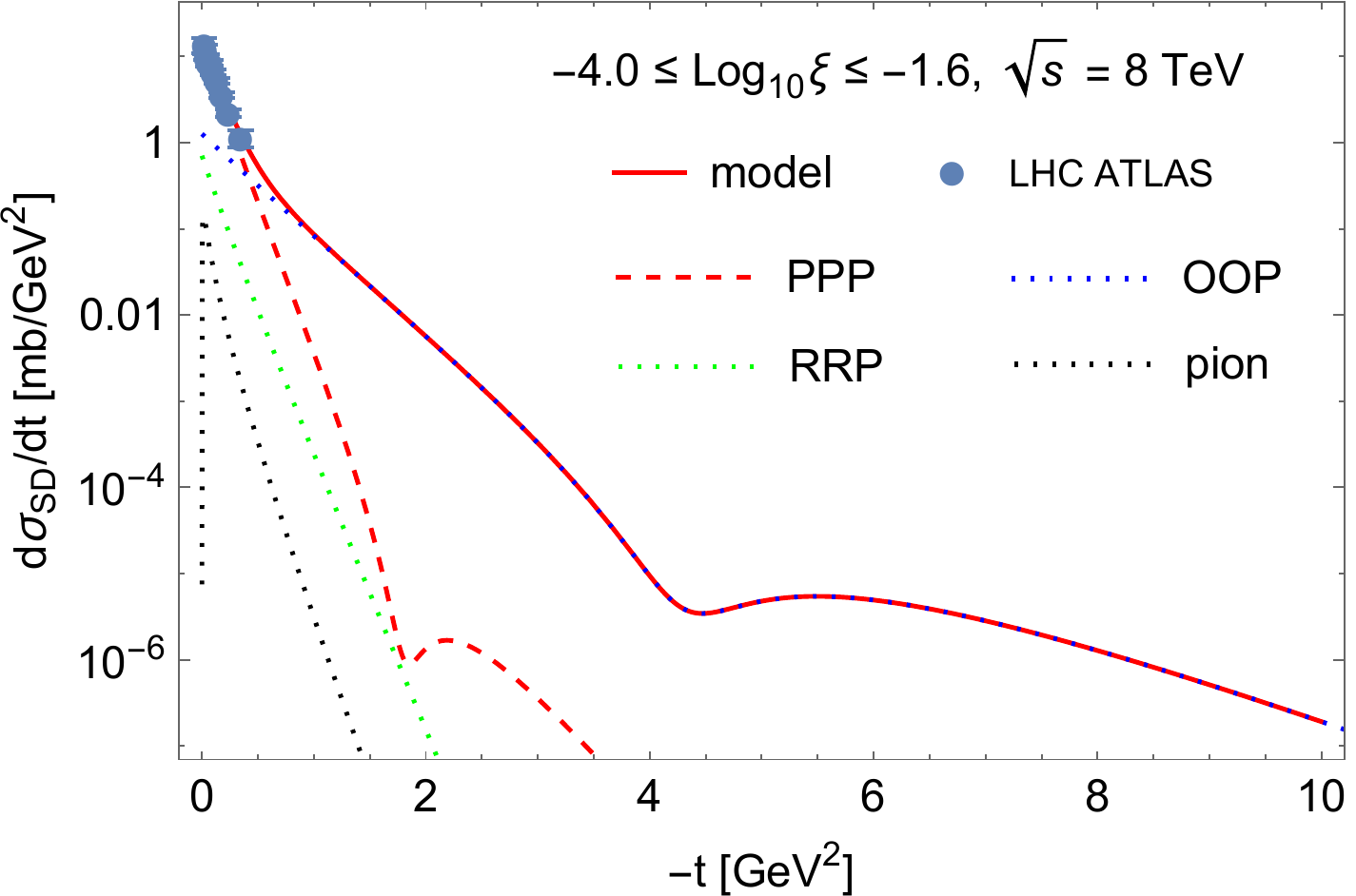}
\caption{Our prediction of the dip-bump structure in the $t$ distribution of the $\xi$ integrated differential cross section of SD at $\sqrt{s}=$ 8 TeV.}
\label{fig:SD_ds_ATLAS_xi_int_t_dep_db}
\end{figure}

Our prediction for the dip-bump structure of the $\xi$ integrated differential cross section of SD at $\sqrt{s}=$ 8 TeV is shown in Figure  \ref{fig:SD_ds_ATLAS_xi_int_t_dep_db}.

\begin{figure}[H]
\includegraphics[width=0.63\textwidth]{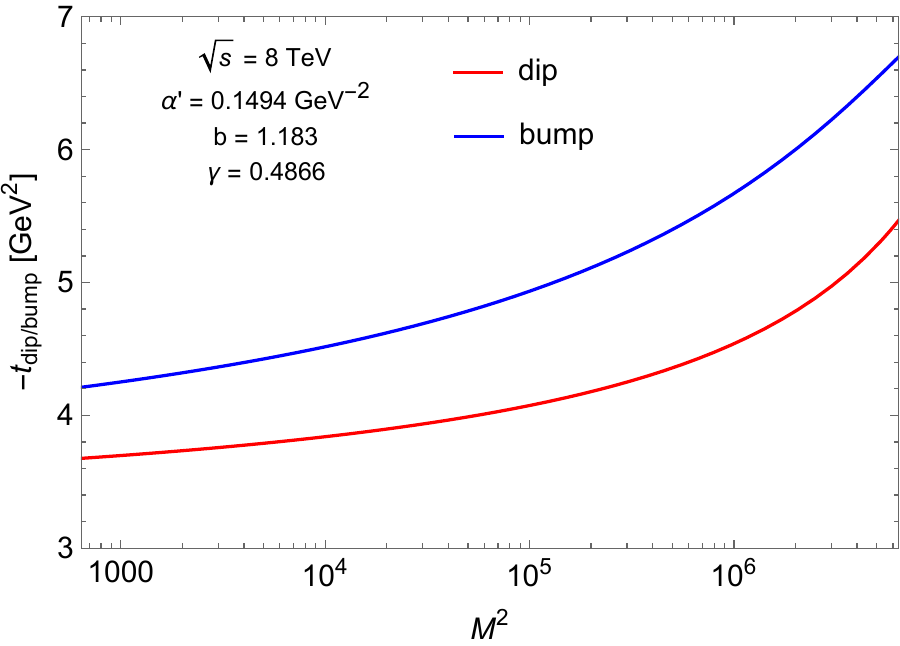}
\caption{Dependence of the position of the dip (red line) and bump (blue line) on the missing mass.}
\label{SD_t_db_M2_dep.pdf}
\end{figure}

Figure~\ref{SD_t_db_M2_dep.pdf} shows the motion of the dip and  bump as functions of $M^2$ at $\sqrt{s}=$ 8 TeV. As $M^2$ increases, the slope of the  $t$ distribution decreases and, consequently, the dip and bump move to higher $|t|$ values.

\vspace{-9pt}

The necessary data sample for measuring the dip-bump structure in SD
at LHC energies in future experiments can be estimated based on the
following assumptions. The differential cross section at the dip position
is on the order of $10^{-13}$ mb/GeV$^4$. Taking bins in missing mass of
$5000$ GeV$^2$ and in $t$ of $\Delta t$ = 0.1 GeV$^2$, the cross section per such a
phase space cell is  about 0.05 pb.

A  measurement with statistical uncertainty better than 10 %
requires a data sample of a minimum of 100 events in the phase space cell
at the dip position. Assuming a factor of 0.2 representing geometrical
acceptance, pile-up rejection and reconstruction efficiency
results in a necessary integrated luminosity of $10^{4}$  pb$^{-1}$.
If the SD events can be measured in both beam directions, then
this integrated luminosity requirement is reduced by a factor of two.

\section{Summary}\label{sec:summary}

In this paper, predictions concerning structures in the differential cross section of single diffraction in the LHC energy region are presented.  Our model was tested on elastic proton--proton data, including the dip-bump structure in the differential cross section, then it was generalized to proton single diffractive dissociation. We have elaborated the model to account for the expected structures in the differential cross section of diffractive dissociation. This is the first prediction of these structures---the existence of a dip and bump. The predictions concern the position of these structures, their dependence on the incoming energy and missing mass. Our predictions (numbers) may slightly change (by a maximum of $20\%$) depending on the results of future measurements, for example, those on a slope, without affecting our main message: a single dip and bump are expected, and their position and motion are predicted.

While structures in elastic $pp$ scattering still remain an important part in studies of high-energy hadron diffraction, a subject of controversy, related studies in SD and DD are missing in the existing literature, probably because of the complexity of the problem. Its understanding is important for further progress in the field.

Our quantitative predictions concerning the structure are based on the relevant values of the SD cross sections and slopes, measured only indirectly or fragmentarily, the missing values/intervals being filled in either by extrapolations or model predictions that make these studies exciting, especially in view of future measurements that will verify the predictions.

To summarize, we conclude the following: 

    (1) \textls[-15]{The Regge-pole model constrained and extended by analyticity, unitarity and duality is the basic tool to handle hadron scattering in the ''soft'' (low and moderate  $t$) region. }
    
     (2) In constructing realistic models, one should not necessarily follow  the ``standard'' procedure based on a simple Regge-pole with a linear trajectory as input. We have presented an alternative approach based on a Regge dipole which can be extended by a non-linear, complex trajectory as~input. 
    
    (3) The present approach enables a smooth, continuous  transition between ``soft'' and ``hard'' dynamics, i.e., from  the ``soft''  exponential to the ``hard'', power-like decrease in the cross sections, obeying quark-counting rules and/or perturbative QCD. In terms of the Regge-pole theory, this transition is mimicked by logarithmic Regge trajectories.
   
    To conclude, a dip followed by a bump is predicted in the $t$ distribution of single diffractive dissociation at $t=-4$ GeV$^2$ in the range of 3 GeV$^2$ $\lesssim|t|\lesssim$ 7 GeV$^2$. 
Apart from the dependence on $s$ (total energy squared) and $t$ (squared momentum transfer), the dependence on missing masses, both at large and low masses in the resonance region, is predicted.  
We are looking forward to future SD measurements at the upgraded LHC and other future experimental facilities.  

\authorcontributions{
Conceptualization, L.J. and I.Sz.; methodology, I.Sz. and R.S.; formal analysis, I.Sz.; writing---original draft preparation, L.J.; writing---review and editing, L.J.,  I. Sz., and R.S.; supervision, L.J.}


\dataavailability{The analyzed data is publicly available (see https://www.hepdata.net/).} 

\acknowledgments{L.J. was supported by EURIZON,
 Grant ID\#38. I.Sz was supported by the NKFIH Grants no. K133046 and K147557, the DOMUS Scholarship of the Hungarian Academy of Sciences, the \'UNKP-23-3 New National Excellence Program of the Ministry for Culture and Innovation from the source of the National Research and MATE KKP 2024. R.S. is supported by the German Federal Ministry of Education and Research under promotion reference 05P21VHK1.}

\conflictsofinterest{The authors declare no conflicts of interest.
} 
\begin{adjustwidth}{-\extralength}{0cm}
\reftitle{References}

\PublishersNote{}
\end{adjustwidth}
\end{document}